\documentstyle[twoside,fleqn,espcrc2,epsfig]{article}
\def\tmt{\times 10^{-2}}
\def\tmth{\times 10^{-3}}
\def\tmf{\times 10^{-4}}
\def\tmfv{\times 10^{-5}}
\newcommand{\beq}{\begin{equation}}
\newcommand{\eeq}{\end{equation}}
\newcommand{\bea}{\begin{eqnarray}}
\newcommand{\eea}{\end{eqnarray}}
\newcommand{\barr}{\begin{array}}
\newcommand{\earr}{\end{array}}
\newcommand{\bc}{\begin{center}}
\newcommand{\ec}{\end{center}}
\newcommand{\btab}{\begin{tabular}}
\newcommand{\etab}{\end{tabular}}
\newcommand{\gv}{\mbox{GeV}}
\newcommand{\tv}{\mbox{TeV}}

\newcommand{\nn}{\nonumber}
\newcommand{\ra}{\rightarrow}

\newcommand{\dro}{\Delta\rho}
\newcommand{\drqcd}{\delta\!\rho_{\rm QCD}}
\newcommand{\roro}{\rho^{(2)}}

\newcommand{\al}{\alpha}

\newcommand{\G}{\Gamma}
\newcommand{\Gmu}{G_{\mu}}
\newcommand{\amu}{a_{\mu}}

\newcommand{\ganu}{\gamma_{\nu}}

\newcommand{\gafi}{\gamma_5}

\newcommand{\noi}{\noindent}

\newcommand{\epm}{e^+e^-}

\newcommand{\sm}{standard model }

\newcommand{\dal}{\Delta\alpha}

\newcommand{\mz}{M_Z^2}
\newcommand{\mw}{M_W^2}

\newcommand{\Dr}{\Delta r}

\newcommand{\alr}{A_{\rm LR}}
\newcommand{\afb}{A_{\rm FB}}

\newcommand{\ass}{asymmetries }

\newcommand{\prd}{Phys.\ Rev.\ D }
\newcommand{\zp}{Z.\ Phys.\ C }
\newcommand{\plb}{Phys.\ Lett.\ B }
\newcommand{\pl}{Phys.\ Lett.\ B }
\newcommand{\prl}{Phys.\ Rev.\ Lett.\ }
\newcommand{\np}{Nucl.\ Phys.\ B }
\newcommand{\elevenbf}{}

\newcommand{\ms}{\overline{MS}}

\newcommand{\AmS}{{\protect\the\textfont2
  A\kern-.1667em\lower.5ex\hbox{M}\kern-.125emS}}

\title{Electroweak interactions: a theoretical overview}

\author{W. Hollik
        \address{Institut f\"ur Theoretische Physik, Universit\"at Karlsruhe\\
        D-76128 Karlsruhe, Germany}}

\begin{document}

\thispagestyle{empty}
\begin{titlepage}
\date{}
\title{
%\hfill {\small hep-ph/}\\
\hfill {\small KA-TP-14-1997} \\[1cm]
{\bf Electroweak interactions: a theoretical overview
} 
}
\author{ {\sc W. Hollik} \\[0.5mm]
%\and 
Institut f\"ur Theoretische Physik\\
Universit\"at Karlsruhe \\
 D-76128 Karlsruhe, Germany \\[5cm]
{\it Plenary talk and } \\
{\it Theoretical Summary of the
     Working Group on Electroweak Interactions} \\
{\it XVI International Workshop on Weak Interactions
      and Neutrinos (WIN '97)} \\
{\it   Capri, Italy } \\
{\it     June 22 - 28,  1997} }

\maketitle

\end{titlepage}

\begin{abstract}
This talk summarizes topical theoretical work for tests of the electroweak
standard model and of the minimal supersymmetric standard model (MSSM).
The status of the standard model 
and the MSSM is discussed in view of recent precision  data.
A brief theoretical summary of the 
Working Group on Electroweak Interactions is included.
\end{abstract}

% typeset front matter (including abstract)
\maketitle

\section{INTRODUCTION}
Impressive experimental results have been obtained for 
the $Z$ boson parameters \cite{moriond}, the $W$ mass
\cite{moriond,wmass}, and the 
top quark mass with
$m_t = 175.6 \pm 5.5$ GeV
\cite{top}

On the other hand, also 
 a sizeable amount of theoretical work has
contributed over the last few years to a steadily rising
improvement of the standard model predictions
(for a review see ref.\ \cite{yb95}). The availability of both 
highly accurate measurements and theoretical predictions, at the level
of nearly 0.1\% precision,
 provides
tests of 
the quantum structure of the standard model thereby
probing its empirically yet untested sector, and simultaneously accesses
alternative scenarios like the minimal supersymmetric extension of  
of the standard model (MSSM).

\section{STATUS OF PRECISION CALCULATIONS}
\subsection{Radiative corrections in the standard model}

The possibility of performing precision tests is based
on the formulation of the \sm as a renormalizable quantum field
theory preserving its predictive power beyond tree level
calculations. With the experimental accuracy 
being sensitive to the loop
induced quantum effects, also the Higgs sector of the \sm
is probed. The higher order terms
induce the sensitivity of electroweak observables
to the top and Higgs mass $m_t, M_H$
and to the strong coupling constant $\al_s$.

Before one can make predictions from the theory,
a set of independent parameters has to be taken from experiment.
For practical calculations the physical input quantities
$ \al, \; \Gmu,\; M_Z,\; m_f,\; M_H; \; \al_s $
are commonly used    
for fixing the free parameters of the standard model.
 Differences between various schemes are formally
of higher order than the one under consideration.
 The study of the
scheme dependence of the perturbative results, after improvement by
resumming the leading terms, allows us to estimate the missing
higher order contributions.
 
\smallskip
Two sizeable effects in the electroweak loops deserve a special
discussion:
\begin{itemize}
\item
The light fermionic content of the subtracted photon vacuum polarization
corresponds to a QED induced shift
in the electromagnetic fine structure constant. The evaluation of the
light quark content
 \cite{eidelman}
 yield the result
\beq  (\dal)_{had} = 0.0280 \pm 0.0007\, . \eeq
Other determinations \cite{swartz}
agree within one standard deviation. Together with the leptonic
content, $\dal$ can
be resummed resulting in an effective fine structure
constant at the $Z$ mass scale:
\beq
   \al(\mz) \, =\, \frac{\al}{1-\dal}\,=\,
   \frac{1}{128.89\pm 0.09} \, .
\eeq
 \item
The electroweak mixing angle is related to the vector boson
masses  by
\bea
  \sin^2\theta  = 
   1-\frac{\mw}{\mz} + \frac{\mw}{\mz} \dro\, +  \cdots
\eea
where
the main contribution to the $\rho$-parameter 
 is from the  $(t,b)$ doublet \cite{rho},
at the present level calculated to
 \beq
 \dro= 3 x_t \cdot [ 1+ x_t \,  \roro+ \drqcd ]
\eeq
with
\beq
 x_t =
 \frac{\Gmu m_t^2}{8\pi^2\sqrt{2}} \, .
\eeq
 The electroweak 2-loop
 part \cite{bij,barbieri} is described by the
function $\roro(M_H/m_t)$.
$\drqcd$ is the QCD correction
to the leading $\Gmu m_t^2$ term
 \cite{djouadi,tarasov}
$$
    \drqcd = - 2.86 a_s  - 14.6 a_s^2, \;\;\;\;
 a_s = \frac{\al_s(m_t)}{\pi} \, .
$$

\end{itemize}
\subsection{The vector boson masses}
The correlation between
the masses $M_W,M_Z$ of the vector bosons,          in terms
of the Fermi constant $\Gmu$, is in 1-loop order given by
 \cite{sirmar}:
\beq
\label{mw}
\frac{\Gmu}{\sqrt{2}}   =
            \frac{\pi\al}{2s_W^2 M_W^2} [
        1+ \Dr(\al,M_W,M_Z,M_H,m_t) ] \, .
\eeq
 
\medskip \noi
The appearance of large terms in $\Dr$ requires the consideration
of higher than 1-loop effects.
At present, the following  
higher order contributions are available:
\begin{itemize}
\item
The leading log resummation \cite{marciano} of $\dal$: \\
$  1+\dal\, \ra \, (1-\dal)^{-1}$
\item
The incorporation of
non-leading higher order terms
containing mass singularities of the type $\al^2\log(M_Z/m_f)$
from the light fermions \cite{nonleading}.
\item
The resummation of the leading $m_t^2$ contribution \cite{chj}
in terms of $\dro$ in Eq.\ (4).
 Moreover, the complete
 $O(\al\al_s)$ corrections to the self energies
 are available \cite{qcd,dispersion1},
 and part of the $O(\al\al_s^2)$ terms \cite{steinhauser}.

\item
The non-leading $\Gmu^2m_t^2 M_Z^2$ contribution 
of the electroweak 2-loop order \cite{padova}.
Meanwhile also the Higgs-dependence of the non-leading
$m_t$-terms has been calculated at two-loop order  
\cite{bauberger} (see also \cite{krause}.
\end{itemize}

\subsection{$Z$ boson observables}
With $M_Z$ as a precise input parameter, 
the predictions for the partial widths
as well as for the asymmetries
can conveniently be calculated in terms of effective neutral
current coupling constants for the various fermions:
\bea
\label{nccoup}
   J_{\nu}^{\rm NC}      &  =  &  
  g_V^f \,\ganu -  g_A^f \,\ganu\gafi     \\ 
  & = &   \left( \rho_f \right)^{1/2}
\left( (I_3^f-2Q_fs_f^2)\ganu-I_3^f\ganu\gafi \right)   \nn
\eea
with form factors 
$\rho_f$ and $s_f^2$ for the overall normalization and the
effective mixing angle.

\smallskip
The effective mixing angles are of particular interest since
they determine the on-resonance asymmetries via the combinations
   \beq
    A_f = \frac{2g_V^f g_A^f}{(g_V^f)^2+(g_A^f)^2}  
\eeq
in the following way:
\beq
\label{afb}
\alr = A_e, \quad  \afb^f = \frac{3}{4}\, A_e A_f \, .
\eeq
Measurements of the \ass hence are sensitive to
the ratios
\beq
  g_V^f/g_A^f = 1 - 2 Q_f s_f^2
\eeq
or to the effective mixing angles, respectively.

\smallskip
The total
$Z$ width $\Gamma_Z$ can be calculated
essentially as the sum over the fermionic partial decay widths.
Expressed in terms of the effective coupling constants they
read up to 2nd order in the fermion masses:
\bea
\Gamma_f
  & = & \G_0
 \, \left(
     (g_V^f)^2  +
     (g_A^f)^2 (1-\frac{6m_f^2}{\mz} )
                           \right)        \nn \\
 &  & \cdot   (1+ Q_f^2\, \frac{3\al}{4\pi} ) 
          + \Delta\G^f_{QCD} \nn
\eea
with
$ \left[ N_C^f = 1
 \mbox{ (leptons)}, \;\; = 3 \mbox{ (quarks)} \right] $ 
\[
\G_0 \, =\,
  N_C^f\,\frac{\sqrt{2}\Gmu M_Z^3}{12\pi},
\]
and the QCD corrections  $ \Delta\G^f_{QCD} $
 for quark final states
 \cite{qcdq}.
The recently obtained non-factorizable part of the 2-loop 
$O(\al\al_s)$ QCD corrections \cite{czarnecki} 
yields an extra negative contribution of 
 -0.59(3) MeV for the total hadronic $Z$ width.

\subsection{Accuracy of the standard model predictions}
 For a discussion of the theoretical reliability
of the \sm predictions one has to consider the various sources
contributing to their
uncertainties:

The experimental error of the hadronic contribution
to $\al(\mz)$, Eq.\ (2), leads to
$\delta M_W = 13$ MeV in the $W$ mass prediction, and
$\delta\sin^2\theta = 0.00023$ common to all of the mixing
angles, which matches with the experimental precision.

The uncertainties from the QCD contributions
can essentially be traced back to
those in the top quark loops for the $\rho$-parameter.
They  can be combined into the following errors
\cite{kniehl95}:
$$
 \delta(\dro) \simeq 1.5\cdot 10^{-4},   \;
 \delta s^2_{\ell} \simeq 0.0001 \, .
$$

The size of unknown higher order contributions can be estimated
by different treatments of non-leading terms
of higher order in the implementation of radiative corrections in
electroweak observables (`options')
and by investigations of the scheme dependence.
Explicit comparisons between the results of 5 different computer codes  
based on  on-shell and $\ms$ calculations
for the $Z$ resonance observables are documented in the ``Electroweak
Working Group Report'' \cite{ewgr} in ref.\ \cite{yb95}.
Table 1  shows the uncertainty in a selected set of
precision observables.
The recently calculated 
non-leading 2-loop corrections
$\sim \Gmu^2m_t^2 M_Z^2$  \cite{padova}
for $\Delta r$ and $s_{\ell}^2$  (not included in table 1)
reduce the uncertainty in $M_W$ and $s^2_{\ell}$ considerably,
by at least a factor 0.5.

\begin{table}[htbp] %\centering
\caption[]
{Largest half-differences among central values $(\Delta_c)$ and among
maximal and minimal predictions $(\Delta_g)$ for $m_t = 175\,\gv$,
$60\,\gv < M_H < 1\,\tv$, $\al_s(\mz) = 0.125$
(from ref.\ \cite{ewgr}) }
\vspace{0.5cm}
\begin{tabular}{@{} c c c}
\hline 
Observable $O$ & $\Delta_c O$  & $\Delta_g O$ \\
\hline
 & & \\
$M_W\,$(GeV)          & $4.5\tmth$ & $1.6\tmt$\\
$\G_e\,$(MeV)          & $1.3\tmt$ & $3.1\tmt$\\
$\G_Z\,$(MeV)          & $0.2$     & $1.4$\\
$ s^2_e$             & $5.5\tmfv$ & $1.4\tmf$\\
$ s^2_b$             & $5.0\tmfv$ & $1.5\tmf$\\
$R_{had}$                 & $4.0\tmth$& $9.0\tmth$\\
$R_b$                 & $6.5\tmfv$ & $1.7\tmf$ \\
$R_c$                 & $2.0\tmfv$& $4.5\tmfv$ \\
$\sigma^{had}_0\,$(nb)    & $7.0\tmth$ & $8.5\tmth$\\
$\afb^l$             & $9.3\tmfv$ & $2.2\tmf$\\
$\afb^b$             & $3.0\tmf$ & $7.4\tmf$ \\
$\afb^c$             & $2.3\tmf$ & $5.7\tmf$ \\
$\alr$                & $4.2\tmf$ & $8.7\tmf$\\
 & & \\
\hline 
\end{tabular}
 
\end{table}
%\normalsize

\section{STANDARD MODEL AND PRECISION DATA}
\begin{figure}[htb]
\vspace{-1cm}
\centerline{
\epsfig{figure=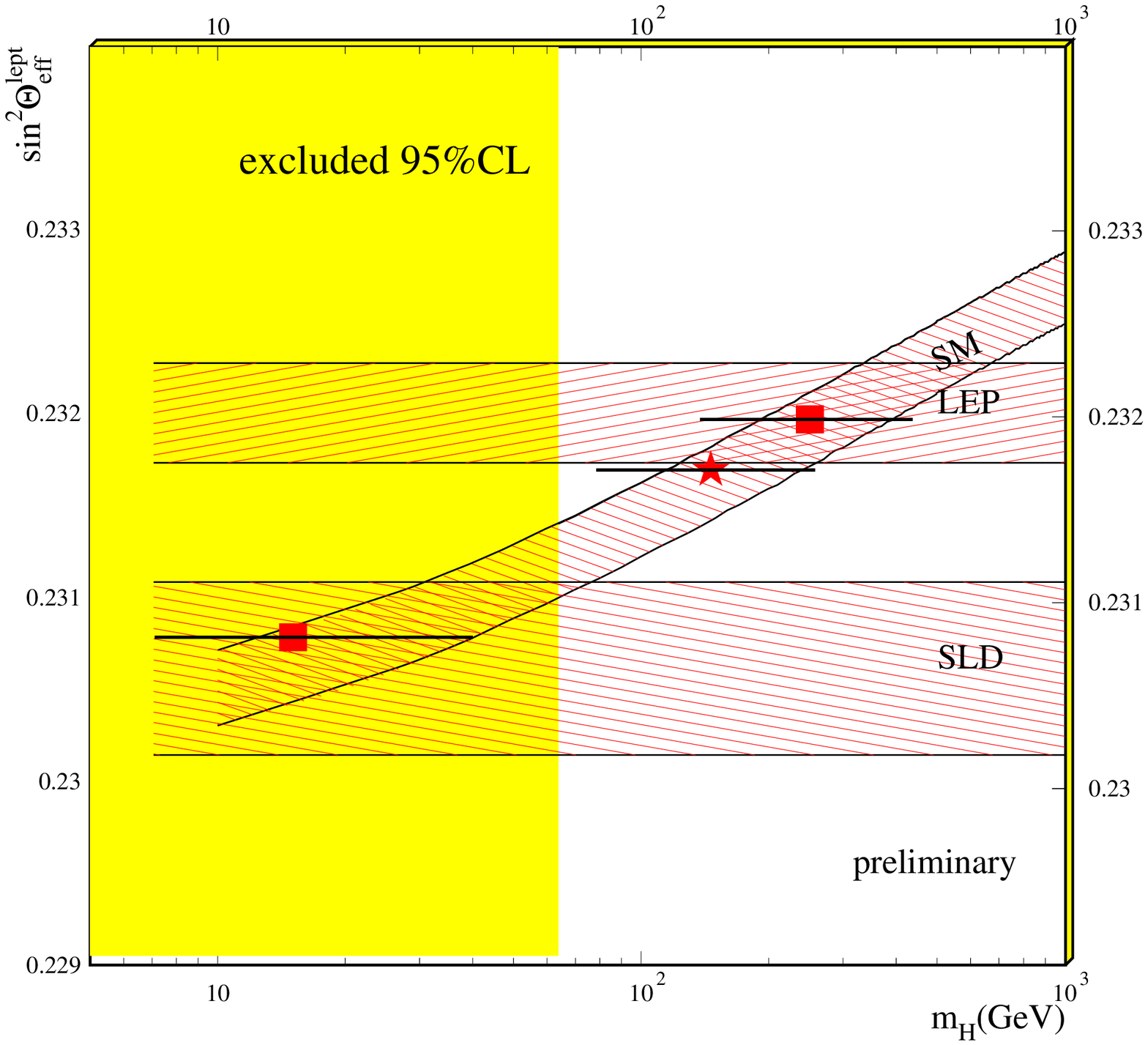,height=8cm}}
\vspace{-1.5cm}
\caption{Dependence of the leptonic mixing angle on the Higgs mass.
         The theoretical predictions correspond to
         $m_t=175\pm 6$ GeV. The SLD 
         and LEP 
         measurements are separately
         shown. The star is the result of a combined fit to LEP and SLD
         data, the squares are for separate fits 
         (from ref.\ {\protect\cite{deboer}}) } 
\label{fig3}
\end{figure}

In this section we put together the standard model predictions for
the discussed set of precision observables for comparison with the most 
recent experimental data \cite{moriond,sld}.
The values for the various forward-backward
asymmetries are for the pure resonance terms (\ref{afb})
only. The small 
photon and interference contributions 
are subtracted from the data,
as well as the QED corrections.
In table \ref{zobs}
the \sm predictions for $Z$ pole observables and the $W$ mass  are
put together for a light and a heavy Higgs particle with $m_t=175$ GeV.
 The last column is the variation of the prediction according to
$\Delta m_t = \pm 6$ GeV. The input value 
for $\al_s$ is chosen as $\al_s = 0.118$ \cite{schmelling}.
Not included are the uncertainties from
$\delta\al_s=0.003$, which amount to 1.6 MeV for the hadronic $Z$ width,
0.038 nb for the hadronic peak cross section, and 0.019 for $R_{had}$.
The other observables are insensitive to small variations of $\al_s$.
The experimental results on the $Z$ observables are from 
LEP and SLD ($A_b, A_c$ and $s_e^2$ from $\alr$).
The leptonic mixing angle determined via $\alr$ by SLD and the 
$s^2_{\ell}$ average 
from LEP differ by about 3 standard deviations:
\bea
   s^2_e(\alr) & = & 0.23055 \pm 0.00041  \nn \\
   s^2_{\ell} ({\rm LEP}) & = & 0.23196 \pm 0.00027 \, . \nn
\eea
The table contains the combine LEP/SLD value.
 $\rho_{\ell}$ and $s^2_{\ell}$ are the leptonic
neutral current couplings in eq.~(\ref{nccoup}), 
derived from partial widths and
asymmetries  under the assumption of lepton universality.
The table illustrates the sensitivity of the various quantities 
to the Higgs mass.
The effective mixing angle turns out to be
the most sensitive  observable, where both the experimental error and the
uncertainty from $m_t$ are small compared to the variation with $M_H$.
Since a light Higgs boson  corresponds to
a low value of $s^2_{\ell}$,
the strongest upper bound on $M_H$ is from $\alr$ at the SLC \cite{sld},
 whereas 
LEP data alone allow to accommodate also a relatively heavy Higgs 
(see figure \ref{fig3}).
Further constraints on $M_H$ are to be expected in the future from
more precise $M_W$ measurements at LEP 2 and the upgraded Tevatron.

\begin{table*}[t]
\vspace*{0.5cm}
\setlength{\tabcolsep}{1.5pc}
            \caption{\label{zobs}Precision observables: 
              experimental results 
             {\protect\cite{moriond}}
             and standard model         
             predictions. }
\begin{center}
 \btab{@{} l  l  r  r  r  }
\hline 
observable & exp.  &  $M_H=65$ GeV & $M_H=1$ TeV &
 $ \Delta m_t $ \\
\hline
 & & & & \\
$M_Z$ (GeV) & $91.1863\pm0.0019$ &  input & input &    \\
%\hline
$\Gamma_Z$ (GeV) & $2.4947\pm 0.0026$ & 2.4974 & 2.4881 & $\pm 0.0015$ \\
%\hline
$\sigma_0^{had}$ (nb) & $41.489\pm 0.055$ & 41.476 & 41.483 & $\pm 0.003$  \\
%\hline
 $ R_{had}$ & $20.783\pm 0.029 $ & 20.753 & 20.725 
                                                     & $\pm 0.002$ \\
%\hline
$R_b$  & $0.2169\pm 0.0009$ & 0.2156 & 0.2157 & $\pm 0.0002$ \\
%\hline
$ R_c$  & $0.1732\pm0.0048$ & 0.1724 & 0.1723 & $\pm 0.0001$ \\
%  \hline
$\afb^{\ell}$ & $0.0177 \pm 0.0010$ & 0.0170 & 0.0144 & $\pm 0.0003$ \\
%\hline 
$\afb^b$ & $0.0979 \pm 0.0022$ &  0.1056 & 0.0970 & $\pm 0.0010$ \\
%\hline
$\afb^c$ & $0.0739 \pm 0.0048$ &  0.0756 & 0.0689 & $\pm 0.0008$ \\
%\hline 
$A_b$            & $0.898\pm 0.050$  & 0.9340 & 0.9350 &  $\pm 0.0001$ \\
%\hline
$A_c$            & $0.649\pm 0.058$  & 0.6696 & 0.6638 &  $\pm 0.0006$ \\
%\hline
$\rho_{\ell}$ & $1.0039\pm 0.0013$ & 1.0056 & 1.0036 & $\pm 0.0006$ \\
%\hline
$s^2_{\ell}$  & $0.23153\pm 0.00022$ & 0.23114 & 0.23264 & $\pm 0.0002$ \\
%\hline
%$s^2_e (A_{LR})$ & $0.23049\pm 0.00050$ & $0.2317\pm 0.0012$   \\
% LEP$+$SLC   &  $0.23143\pm 0.00028$    &                    \\
%\hline
$M_W$ (GeV) & $80.43 \pm 0.08$ & 80.417 & 80.219 & $\pm 0.038$  \\
  & & & & \\
\hline
\etab
 \vspace{0.5cm}
\end{center}
%  \vspace{-1.5cm}
\clearpage
\end{table*}

\smallskip 
Note that
the experimental value for $\rho_{\ell}$ points out  the presence of
genuine electroweak corrections by 3 standard deviations. The 
deviation from the \sm prediction in the quantity $R_b$ has been reduced
to one standard deviation by now.
% (see Figure \ref{rbfig}).
Other small deviations 
 are observed in the asymmetries: the purely leptonic $\afb$ is slightly
higher than the standard model predictions, and $\afb$ for $b$ quarks is 
lower.
Whereas the leptonic $\afb$ favors a very light Higgs boson, 
the $b$ quark asymmetry needs a heavy Higgs. The measured asymmetries deviate
from the best fit  values by 2-3 standard deviations.

\smallskip 
The $W$ mass prediction in table \ref{zobs}
is obtained  from Eq.~(\ref{mw}) 
(including the higher order terms)
from
 $M_Z,\Gmu,\al$ and  $M_H,m_t$.
The present experimental value for the $W$ mass
from the combined UA2,  CDF and D0 results \cite{wmass} is
\beq
\label{mwpp}
  M_W^{\rm exp}\, = \, 80.41 \pm 0.09 \, \gv \ , 
\eeq
and from LEP 2 \cite{moriond}:
\beq
\label{mwlep}
  M_W^{\rm exp}\, = \, 80.48 \pm 0.14 \, \gv \ , 
\eeq
yielding the average given in table \ref{zobs}.

\smallskip
The quantity $s_W^2$ resp.\  the ratio $M_W/M_Z$
can indirectly be measured in deep-inelastic
neutrino-nucleon scattering.
The present world average on $s_W^2$ from the experiments
CCFR, CDHS and CHARM \cite{neutrino} 
\beq
\label{sw}
s_W^2 = 1 - M_W^2/M_Z^2 = 0.2236 \pm 0.0041   
\eeq
corresponds to $M_W = 80.35 \pm 0.21$ GeV and hence 
is fully consistent with the direct vector boson mass measurements
and with the standard theory. 

\smallskip 
The indirect determination of the       
$W$ mass  from the global fit to the LEP1/SLD data \cite{moriond}
\[ M_W = 80.323\pm 0.042 \, \gv \, , \]
is slightly lower, but still
in agreement with the direct measurement.

\bigskip \noi
{\it Standard model global fits:} 

\smallskip 
In the meantime the data have reached an accuracy such that
global fits with 
respect to both $m_t$ and $M_H$ as free parameters have become 
available \cite{moriond,haidt,higgsfits,deboer}. 
The results of ref.~\cite{moriond} based on the
most recent  LEP (still preliminary) and SLD data
are
\[ m_t = 156^{+11}_{-9}\, \gv, \quad 
   M_H = 39^{+63}_{-20} \, \gv \, .
\]
Without the
Tevatron constraint on the top mass, the favored range for $m_t$ is
a bit lower than the direct measurement. The reason for this
behaviour is the strong impact on the
upper limit of $m_t$ from the quantity $R_b$.
%(see Figure \ref{rbfig}).

Together with the top mass as an additional experimental data point,
the global fit to all electroweak results from
LEP, SLD, $M_W$,  $\nu N$ and $m_t$ 
yields  the following results \cite{moriond} 
for $m_t$ and $\al_s$
\[ m_t = 173.3 \pm 5.4 \, \gv, \quad 
   \al_s = 0.120 \pm 0.003 \]
and for the Higgs mass
\[ M_H = 121^{+119}_{-68} \, \gv  \]
with an overall  $\chi^2 = 21/15$.
The input from $\alr$
is decisive for a restrictive upper bound for $M_H$. 
Without $\alr$,  the 95\% C.L upper bound is shifted upwards by about
260 GeV \cite{deboer}.
The value obtained for
 $\al_s$ is in very good agreement with the world average
\cite{schmelling}.

%\clearpage

\smallskip \noi
The numbers given above 
 do not yet include the theoretical uncertainties of the
standard model predictions. The LEP Electroweak Working Group
 \cite{moriond}
has performed a study of the influence of the various `options'
discussed in section 2.4 on the bounds for the Higgs mass with the result
that the 95\% C.L. upper bound is shifted by nearly 
100 GeV to higher values, yielding
\[ M_H < 430 \, \gv \, (95\% \, {\rm C.L.}) \, . \]
It has to be kept in mind, however, that this error estimate is based on the
uncertainties as given in table 1. Since the recent improvement in the 
theo\-retical prediction \cite{padova}
is going to reduce the theo\-retical uncertainty
one may expect also a significant
smaller theoretical error on the Higgs mass bounds once the 2-loop terms
$\sim \Gmu^2 m_t^2 \mz$ are implemented in the codes used for the fits.
At the present stage the analysis is done without the new terms.

\smallskip 
The error from  the hadronic vacuum polarization is 
incorporated in the fit and is thus part of the result on the Higgs mass
bound. The uncertainty induced from $\dal$ is 
quite remarkable at the present stage (see for example the discussion
in \cite{haidt}).

\medskip
There are also a theoretical constraints on the Higgs mass
from vacuum stability and absence of a Landau pole \cite{lindner},
and from lattice calculations \cite{lattice}. Recent calculations
of the decay width for $H\ra W^+W^-,ZZ$  in the large $M_H$ limit
in 2-loop order \cite{ghinculov} have  shown that the 2-loop
contribution exceeds the 1-loop term in size (same sign) for
 $M_H > 930$ GeV. The requirement of applicability of
perturbation theory therefore puts a stringent upper limit on the
Higgs mass. The indirect Higgs mass bounds obtained from the
precision analysis show, however, that the Higgs boson is well below
the mass range where the Higgs sector becomes non-perturbative.

\section{$W$ BOSONS IN $\epm$ COLLISIONS}
At LEP 2, pair production of on-shell
$W$ bosons can be studied experimentally allowing 
$M_W$ measurements with an aimed error of about 40 MeV and tests
of the trilinear vector boson self couplings.
For this purpose 
standard model calculations for the process
$\epm \ra W^+W^- \ra 4 f$ and the corresponding 4-fermion background
processes  are required 
at the accuracy level of 1\%.
For practical purposes, improved Born approximations are in use for
both resonating and non-resonating processes, dressed by initial state
QED corrections, incorporating the set of fermion loop contributions at
the one-loop level in the double- and single-resonating   
processes (see \cite{ww} and talk by Passarino \cite{giampiero} in these
proceedings).

\section{ELECTROMAGNETIC DIPOLE MOMENTS} 
\subsection{Muon anomalous magnetic moment}
The anomalous magnetic moment of the muon,
\beq
   a_{\mu} = \frac{g_{\mu}-2}{2}
\eeq
provides a precision test of the standard model at low energies.
Within the present experimental accuracy of
$\Delta\amu = 840\cdot 10^{-11}$, theory and experiment are in best
agreement, but the electroweak loop corrections are still hidden
in the noise. The new experiment E 821 at Brookhaven National
Laboratory \ is being prepared  to reduce
the experimental error down to $40\pm 10^{-11}$ and hence will
become sensitive to the electroweak loop contribution.

For this reason the standard model prediction has to be known with
comparable precision. Recent theoretical work has contributed to
reduce the theoretical uncertainty by calculating the electroweak
2-loop terms \cite{ew22,ew23}
and updating the contribution from the hadronic
photonic vacuum polarization
(first reference of \cite{eidelman})
\[
 \amu^{had}(\mbox{vacuum pol.}) = (7024\pm 153)\cdot 10^{-11}
\]
which agrees within the error with the result of \cite{dub}.
The
main sources for the theoretical error at present are the hadronic
vacuum polarization and the light-by-light scattering mediated by
quarks, as part of the 3-loop hadronic contribution
\cite{sanda,bijnens}.
Table 3 contains  the breakdown of $\amu$. The hadronic part
is supplemented by the higher order $\al^3$
vacuum polarization effects \cite{had3} but is without the
light-by-light contribution (see also \cite{krause}).

\begin{table}[htbp] %\centering
\caption[]
{Contributions $\Delta\amu$ to the muonic anomalous magnetic moment
and their theoretical uncertainties, in units of $10^{-11}$.  }
\vspace{0.5cm}
\begin{tabular}{@{} l r r}
\hline 
source & $\Delta\amu$  & error \\
\hline
 &  &  \\
QED \cite{qed} & 116584706 & 2  \\
hadronic \cite{eidelman,had3}  & 6916  & 153 \\
EW, 1-loop \cite{ew1} & 195 &     \\
EW, 2-loop \cite{ew23}  & -44   &  4  \\
light-by-light \cite{sanda} & -52 &  18 \\
light-by-light \cite{bijnens} & -92 &  32 \\
  &  &  \\
\hline
future experiment &   & 40  \\
\hline 
\end{tabular}
 
%\label{ta9}
\end{table}
%\normalsize
 
The 2-loop electroweak contribution is as big in
size as the expected experimental error.
The dominating theoretical uncertainty at present is the error in
the hadronic vacuum polarization. But also the contribution involving
light-by-light scattering needs improvement in order to reduce the
theoretical error.

\subsection{Electric dipole moments}
Electric dipole moments of the fundmantal fermions are 
CP violating quantitites. In the standard model they are introduced
via the complex phase in the CKM matrix at the three-loop level and
hence are very small quantitites. A recent new standard model
calculation of the
electric dipole moment of the neutron \cite{neutron,krause}
composed from the dipole
moments of the constituent quarks yields the value
 \beq
 d_{\rm n} \simeq  10^{-34} \, e\, {\rm cm} 
\eeq
which is several orders of magnitude below the experimental limit
$ \mid d_{\rm n}^{\rm exp} \mid < 10^{-25} \, e\, {\rm cm}$
\cite{rev96}.
An experimental observation of a non-zero electric dipole moment would
therefore be a clear indication for the presence of 
non-standard physics.

\subsection{Electric dipole form factors of the top quark}
In Supersymmetry, CP violating electric dipole form factors can already
be induced at the one-loop level by complex values of the SUSY
parameters, yielding complex coupling constants. 
Since the effects are enhanced for large fermion masses, the 
most sizeable deviations from the small  standard model predictions 
would occur for top quarks. 
A study  in the
minimal supersymmetric standard model (MSSM) \cite{bartl} has shown that
the contribution to CP violating observables in top production and decay
processes can reach the level of a few $10^{-3}$ and thus should 
in principle be observable at future $\epm$ colliders.

\section{VIRTUAL EFFECTS FROM SUPERSYMMETRY}
\subsection{Precision tests of the MSSM} 
The MSSM is presently 
the most predictive framework beyond the standard model.
Its structure allows a similarly complete calculation of
the electroweak precision observables
as in the standard model in terms of one Higgs mass
(usually taken as $M_A$) and $\tan\beta= v_2/v_1$,
together with the set of
SUSY soft breaking parameters fixing the chargino/neutralino and
scalar fermion sectors.
It has been known since quite some time
\cite{higgs}
that light non-standard
Higgs bosons as well as light stop and charginos
% all around 50 GeV or little higher,
predict larger values for the ratio $R_b$ \cite{susy1,susy3}.
Complete 1-loop calculations are available for
$\Delta r$ \cite{susydelr} and for the $Z$ boson observables
\cite{susy3}.

\smallskip
A possible mass splitting between
 $\tilde{b}_L$-$\tilde{t}_L$  yields a contribution to the $\rho$-parameter
of the same sign as the standard top term.
As a universal loop contribution, it enters the quantity $\Dr$ and the
$Z$ boson couplings
and is thus significantly constrained by the data on $M_W$ and the
leptonic widths. Recently the 2-loop $\al_s$ corrections have been
computed \cite{drosusy}
which can amount to 30\% of the 1-loop
$\Delta \rho_{\tilde{b}\tilde{t}}$.

\smallskip
The MSSM yields a similar good description of the precision data as
the standard model.
A global fit 
\cite{deboer,schwick} 
to all electroweak precision data, including the top mass measurement,
shows that
 the $\chi^2$ of the fit is slightly better than in the
standard model, but due to the larger numbers of parameters, the
probability is about the same as for the 
Standard Model (see \cite{schwick}).

\begin{figure}[htb]
\vspace{-2cm}
\centerline{
\mbox{\epsfxsize8.0cm\epsffile{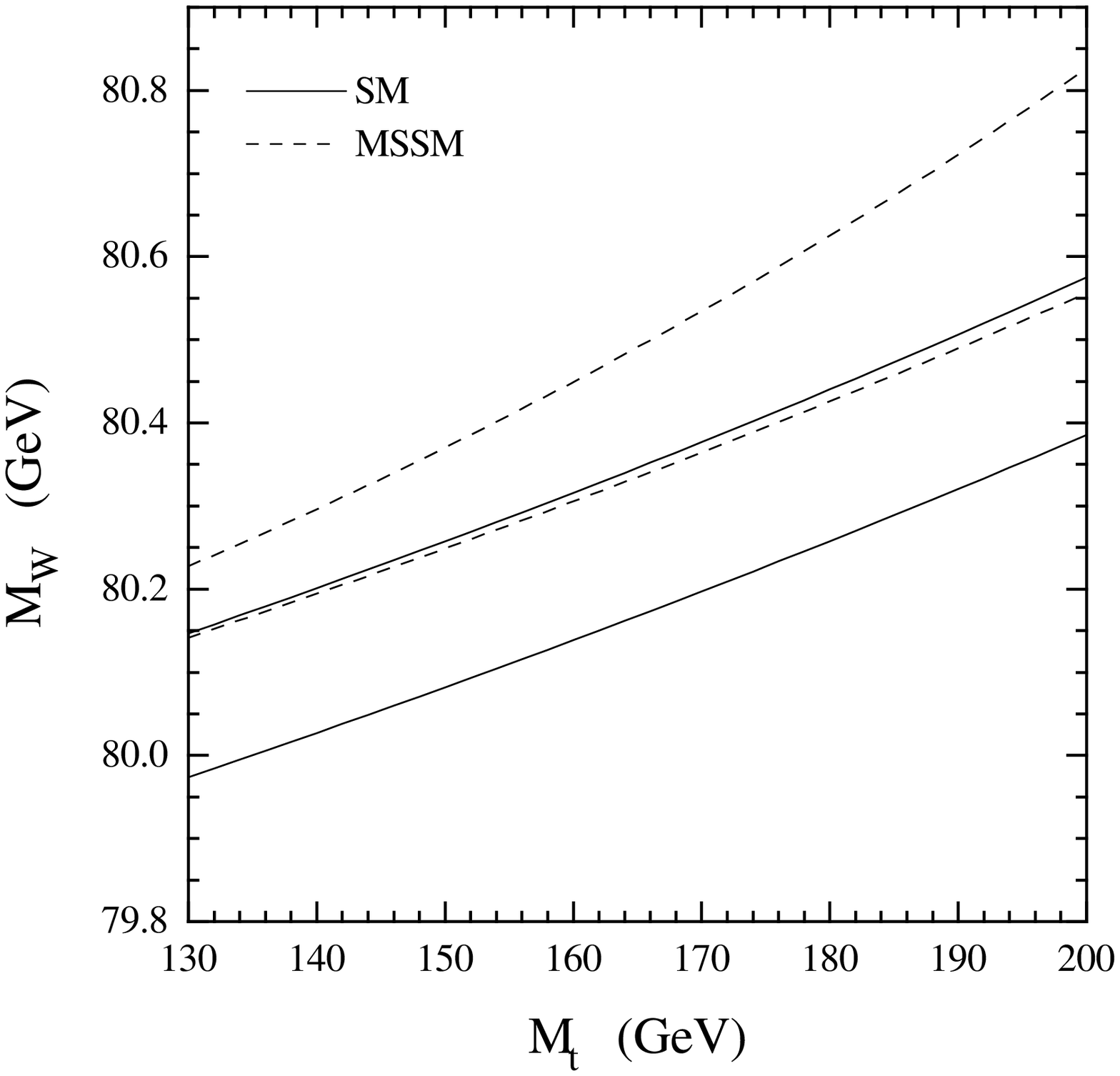}}} 
\vspace{-3cm}
\caption{The $W$ mass range in the standard model (-----) and the
         MSSM (- - -). Bounds are from the non-observation of Higgs
         bosons and SUSY particles at LEP2.} 
\label{susymw}
\end{figure}

\bigskip
Figure \ref{susymw}
displays the range of predictions for $M_W$ in the minimal model
and in the MSSM. Thereby it is assumed that no direct discovery has been
made at LEP2. As one can see, precise determinations of $M_W$ and $m_t$
can become decisive for the separation between the  models.
A large value of $M_W$, as presently experimentally observed,
gives preference to the MSSM.

\setlength{\unitlength}{0.7mm}
%\clearpage

\subsection{Quantum SUSY signatures in top decays} 
In the supersymmetric version of the standard model
additional top decay modes into SUSY particles
$t \ra \tilde{t} \chi^0,\, \tilde{b} \chi^+$ and into
charged Higgs bosons $t \ra b H^+$ are possible in certain
regions of the parameter space. These channels may affect the
branching ratio for the standard top decay
$t \ra b W^+$ in a sizeable way and hence have to be treated
carefully for an overall and systematic discussion of the MSSM.
For the required accuracy, quantum corrections to the two-body 
decays have to be taken into account as well as the next-to-leading
3-particle decay channnels. Whereas the loop corrections to the
standard top decay \cite{top1}
are not very significant (below 10\%), the 
supersymmetric QCD corrections to the non-standard decays are
remarkable \cite{top2,sola}.
In particular for the decay mode $t\ra b H^+$ they reach
30-50\% for large values of $\tan\beta$ and compensate the large
standard QCD corrections. The electroweak two-loop terms
are also large, up to 50\%, such that in the overall sum
significant quantum corrections arise with the
sign opposite  
to the conventional QCD corrections. This typical quantum
signature may be considered as an imprint of virtual supersymmetry
which remains sizeable enough even in the absence of any direct top decay
into genuine SUSY particles. For a more detailed presentation
see the talk by Sol\`a in these proceedings \cite{sola}.

\section{CONCLUSIONS}
The experimental data for tests of the standard model 
have achieved an impressive accuracy.
In the meantime, many 
theoretical contributions have become available to improve and 
stabilize the \sm predictions. To reach, however, a theoretical
accuracy at the level of 0.1\% or below, new experimental data on
$\dal$ and more complete electroweak 2-loop calculations are required.

Also impressive is the overall agreement between
theory and experiment for the entire  set of the
precision observables with a light Higgs boson in the non-perturbative
regime, although a few observed deviations from the
standard model predictions reduce the quality of the global fit.

The MSSM, mainly theoretically advocated,
is competitive to the standard model in describing the data
with about the same quality in global fits.
Outside the frame of the MSSM, supersymmetry with broken $R$-parity
provides
a viable scenario to accommodate leptoquarks which can account for the
observed anomaly in deep-inelastic scattering processes at HERA at high
momentum transfer (see talk by Spiesberger in these proceedings
\cite{spiesberger}).

\medskip \noi
{\bf Acknowledgements:}
I want to thank A. Bartl, S. Heinemeyer, B. Krause, K. Moenig, 
G. Passarino,
U. Schwickerath, H. Spiesberger
and G. Weiglein for helpful discussions and valuable
informations.

\end{document}